\documentclass[10pt]{article}  
\usepackage{amssymb,amsmath, cite,epsfig, psfrag, ulem, color} 
\newcommand{\be}{\begin{equation}}  
\newcommand{\ee}{\end{equation}}  
\newcommand{\bea}{\begin{eqnarray}}  
\newcommand{\eea}{\end{eqnarray}}  
\newcommand{\bean}{\begin{eqnarray*}}  
\newcommand{\eean}{\end{eqnarray*}}

\newcommand{\gapproxeq}{\lower  
.7ex\hbox{$\;\stackrel{\textstyle >}{\sim}\;$}}  
\newcommand{\lapproxeq}{\lower  
.7ex\hbox{$\;\stackrel{\textstyle <}{\sim}\;$}}

\newcommand{\bc}{\begin{center}}  
\newcommand{\ec}{\end{center}}  
\newcommand{\btab}{\begin{tabular}}  
\newcommand{\etab}{\end{tabular}}

\def\qq{$ q\bar q $}  
\def\ss{$ s\bar s $}

\def\dd{$d\bar d$}  

\def\10bar{$\mathbf{\overline{10}}$}

\def\3bar{$\bar {\hbox{\bf 3}}$} 
\def\Ts{$ud\bar{s}$} 
\def\TQ{$ud\bar{Q}$}

\def\udss{$ud\bar{s}\bar{s}$}

\begin{document}  
\normalem  
\begin{titlepage}  
  
  \baselineskip=18pt \vskip 0.9in
\begin{center}  
  {\bf \Large Pentaquark implications for Exotic mesons}\\
  \vspace*{0.3in} {\large T.Burns}\footnote{\tt{e-mail:
      burns@thphys.ox.ac.uk}}
  {\large and F.E.Close}\footnote{\tt{e-mail:f.close@physics.ox.ac.uk}}\\
  \vspace{.1in} {\it Peierls Centre for Theoretical Physics, University of Oxford,\\
    1 Keble Rd., Oxford OX1 3NP, UK} \\
  \vspace{0.1in}
  {\large and J.J.Dudek}\footnote{\tt{e-mail:dudek@jlab.org}}\\
  \vspace{.1in}
  {\it Thomas Jefferson National
          Accelerator Facility, Newport News, VA 23606, USA}\\

\end{center}  
  
\begin{abstract}  
  If the exotic baryon $\Theta^+(1540)$ is a correlated $udud\bar{s}$
  with $J^P = \frac{1}{2}^+$, then there should exist an exotic meson,
  $J^P =  1^-$ $\vartheta^+ (S=+2)$ $\to K^+K^0$ $\sim 1.6$GeV with width $O(10-100)$MeV.  The
  $\pi_1(1400;1600)$ may be broad members of {\bf 10} $\pm$ \10bar in
  such a picture. Vector mesons in the 1.4 - 1.7GeV mass range are also compared with
  this picture.

\end{abstract}  
\end{titlepage}  
 

\subsection*{Introduction}

If the narrow $\Theta^+$ seen in $nK^+$ and $pK^0$ is confirmed as a pentaquark, then
correlations among quarks in Strong QCD play an essential role. There
is considerable literature recognising that $ud$ in colour \3bar
with net spin 0 feel a strong attraction\cite{scalardiq,dgg}.  We
denote this $([ud]^{\bar{3}_c}_0)$, the subscript denoting its spin,
the superscript being the colour and the ( ) denoting the 
quasiparticle.  Ref.\cite{jw} considers the following subcluster for
the pentaquark: $([ud]_0)([ud]_0)\bar{s}$ with a $P$-wave between the
assumed bosonic $(ud)$ correlations.  By contrast ref.\cite{kl1}
assumes that the $([ud]_0)$ seed is attracted in $P$-wave to a
strongly-bound ``triquark" $([ud]^{6_c}_1\bar{s})$. These models
assign the $\Theta^+$ to \10bar of flavour.

An essential feature of these dynamics is that in $S$-wave the chromomagnetic
repulsion of like flavours destabilises the configuration such that decay
to meson + baryon in $S$-wave has such a large width as to be effectively 
nonexistent\cite{jafexotic}. It is in the $P$-wave that potentially interesting
pentaquarks emerge.

Mixing between the two configurations $([ud]^{\bar{3}_c}_0)\bar{s}$ 
and $([ud]^{6_c}_1\bar{s})$ has been shown to lead to an eigenstate of low mass,
which may be identified with the $\Theta(1540)$\cite{maltman,hog,vento}.
Further, this mixing potentially stabilises the $ud\bar{s}$ configuration,
underpinning the metstability of the $\Theta$\cite{vento}.
  
The point of departure for this paper is to note that if either of
these correlations is realised empirically for the $P$-wave, then on model independent
grounds one can replace $([ud]^{\bar{3}_c}_0)$ by $\bar{q}$, which
implies the existence of \10bar and {\bf 10} exotic mesons. A specific
example is the analogue of $\Theta^+ \equiv$ (\Ts)$(ud)$ $\to$
$\vartheta^+ \equiv $ (\Ts)$\bar{s}$.  
While many of these may be broad and unmeasurable, we shall suggest
that if the mixing that lowered and stabilised the $\Theta$ configuration in ref\cite{vento}
applies, then this leads to an observable $J^P = 1^-$ $\vartheta$ with strangeness
=+2, together with other $J^{PC} = 1^{-\pm}$ states with rather characteristic signatures.
  If the $\Theta^+$ should survive
high statistics data and with a width of $\sim 1$MeV, then the
observation or otherwise of such mesons may help to discriminate among
models for the dynamical origin of that metastability. 
Recent proposals \cite{vento} to explain the anomalously narrow 
width of the $\Theta^+$ ought to carry over to the meson world, 
although due to greater phase space we expect that the meson analogue
 $\vartheta^+$ will have width of order $10-100$MeV. Analogues
 with $J^P = 0^-,2^-,3^-$ also arise but are expected to be broad and
 unobservable.
 
 The possibility that the $\pi_1(1400)$ and $\pi_1(1600)$\cite{pdg04} could belong
 to these multiplets is discussed. In addition, the pattern of vector mesons in the
 1.4 - 1.7GeV mass region\cite{pdg04} may also have some overlap with these ideas.
 
 \subsection*{Exotic Mesons}

The idea that mesons beyond \qq~ exist is not new. Jaffe\cite{jaf77}
noted that the attractive forces alluded to above lead to a low lying
$qq\bar{q}\bar{q}$ $S$-wave  nonet. There is good evidence that
the scalar mesons below 1 GeV are intimately related to such a
picture\cite{ct02} and the idea of these as correlated diquarks has been
resurrected by ref\cite{maiani}. Models of multiquark mesons typically predict a 
large number of
  states, in various flavour multiplets and spin states. This usually
  includes exotics, whose flavour or $J^{PC}$ cannot be
  constructed from $q\bar{q}$. Any model of exotics must face the fact that
   there is at most only a handful
  of such candidates. 
  The full set of flavour representations from 
  $\mathbf{3\otimes 3\otimes \bar{3}\otimes\bar{3}=(\bar{3}+6)\otimes(3+\bar{6})}$ is:
  \\
(i) $\mathbf{\bar{3}\otimes 3=8\oplus 1}$, Jaffe's original nonet\\
(ii)$(\mathbf{6\otimes 3) \oplus (\bar{3}\otimes\bar{6})
=10\oplus 8_a \oplus\overline{10}\oplus 8_b
=18\oplus \overline{18}}$, the decuplets into which ref.\cite{chung} assign the 
$\pi_1(1400)$ state\cite{pdg04}, and accompanying octets, and\\
(iii) $\mathbf{6\otimes\bar{6}=27+8+1}$

All such states were included in the original study of
$[qq][\bar{q}\bar{q}]$ in $S$-wave\cite{jaf77}.  Within the dynamical
assumptions made there, all states other than the nonet (i) were predominantly expected to be
very broad and effectively unobservable. With $L=1$ in the system there are many multiplets with 
negative parity,
$J=0,1,2,3$ and $C=\pm$.  Either all the states are broad and
unobservable or some organising principle is required if one wishes to
identify one or two states as members, as e.g.\cite{chung}, and explain away the remainder. 
We examine the  $qq\bar{q}\bar{q}$ systems in $L=1$  in light of
 the recent interest in diquark correlations, focussing on the supposed dominance
 of ``good" diquarks over ``bad" \cite{jafexotic}. We shall see that the
 correlation of triquark-antiquark (and charge conjugate) leads to a limited spectrum of $J^{PC}$ 
states, in particular suggesting   a  lowering of the $1^{-\pm}$ multiplets
 to mass scales akin to those of the $\pi_1(1400;1600)$ claimed in  
 $\pi \eta$ and $\pi\eta'$ 
respectively\footnote{We refer the reader to \cite{indiana} for an analysis of the experimental data that disputes a resonant interpretation for the $\pi_1(1400)$}\cite{pdg04}.

\subsection*{The $\vartheta^+$: an exotic meson analogue of the $\Theta^+$}

If the $\Theta^+$ is
confirmed  as a resonant state with $\Gamma \sim 1$MeV,
then the stability of the correlations, at least in $P$-wave, raises
interesting questions for the existence of observable meson analogues in 
representations {\bf 18} and $\overline{\bf 18}$.
In the (\TQ)$\bar{q}$ correlation, at least, the same
dynamics that lead to  $\Theta^+$ in \10bar imply a \10bar of mesons
which will include a  $\vartheta^+$ with strangeness $+2$,
and its {\bf 10} antiparticle with strangeness $-2$. In a later section we will show that a 
dynamical picture in which a triquark-antiquark (and charge conjugate)  are in $L=1$ suggests that 
the $1^{-(\pm)}$  multiplet lies lowest. 

 Neither of the correlations of \cite{jw} or \cite{kl1} alone derives
 either the low mass or width of the $\Theta^+(1540)$
 readily\cite{fc04ichep}. The triquark correlation in ref\cite{kl1}, with the $[ud]$ in the
 $(ud\bar{s})$ being in the configuration $[ud]^6_1$ was
 chosen for its maximal attraction. However, it has been widely noted
 \cite{hog,vento,maltman} that mixing with the $[ud]^{\bar{3}}_0$ via either
 one-gluon exchange, instanton forces or the effect of virtual $KN$ loops, leads to 
 one eigenstate that
 is lower in energy than either of the unmixed states. That this could cause a decoupling of the 
$\Theta^+ \to KN$, along the
 lines suggested in ref\cite{klmix}, is possible but has not been
 demonstrated (and  this mechanism has problems if virtual $K^* N$ loops are included). 

The energy levels in such a situation are interesting.
 Ref\cite{kl1} consider the \TQ $ $ correlation to be $([ud]^6_1\bar{s})$
 for which the chromomagnetic forces are highly attractive, the attractive interaction 
 between  $[ud]^6_1$ diquark and antiquark compensating for the reduced diquark attraction 
 relative to  $[ud]^{\bar{3}}_0$. The
 correlation $([ud]^{\bar{3}}_0\bar{s})$, where there is no such compensating diquark-antiquark 
 hyperfine interaction, is relatively disfavoured.  However,
 the same chromomagnetic forces cause a mixing between these two
 configurations and the emergence of a light and heavy eigenstate.
 Ref\cite{vento} argues that including both gluon exchange and
 instanton forces in the mixing analysis, leads to effective masses
 for the light eigenstate \Ts $ \sim 750$MeV and $m[ud] \sim 450$MeV.
 The reason for this being some 350 MeV more bound than the lightest
 $uds$ $\Lambda$ state is that the gluon and instanton forces are
 twice as attractive in the \qq~ channel than in the $qq$ case.
 Further, as m(\Ts)$ \leq m(K)+m(d/u)$, for constituent quark mass $m(q)$, the pentaquark cannot
 dissociate into the $Ku(d)$. It is proposed\cite{vento} that
 rearrangement is suppressed in $L=1$ with the result that the $\Theta^+$
 is metastable.

 The reduced mass of the 750 MeV triquark and 450 MeV diquark is 280
 MeV. The similarity of this to the strange quark in the $\phi$, which
 is $\sim m(\phi)/4 =255$MeV, suggests a similar price for
 $L$-excitation in the two systems. Using $m(f_2(1525)/f_1(1420)) -
 m(\phi) \sim 400- 500$MeV as a measure of the orbital excitation
 energy\footnote{ Note that this is more reasonable than the $207$MeV claimed
  by KL \cite{kl1} 
 on the basis of an analogy to the $D_s$ spectrum, see \cite{jjdFermilab} 
 for a critique.} giving the mass scale for $\Theta^+$ $\sim 1600-1700$MeV. Spin
 orbit splitting might reduce this to 1540MeV\cite{fcdudekls}.

But now consider the $[ud]$ accompanying the \Ts $ $ in the $P$-wave: we can replace this
 by any of $\bar{u}, \bar{d}$ or $\bar{s}$ to form a meson. Were we to
 do so for any combination $q_iq_j\bar{q_k}$ and $\bar{q_l}$, we would
 have a {\bf 10} and \10bar of mesons with a $P$-wave internally. 
 The $\pi_1(1400;1600)$ could be
 members of such a multiplet (this was originally proposed on symmetry
 grounds in ref\cite{chung}); however, an inescapable consequence of
 such a proposal is that there exists an exotic \Ts$\bar{s}$ meson with
 strangeness +2. With $m(\bar{s}) \sim m[ud]$ we predict this to be
 at $\sim 1600$MeV. Assuming that the $\Gamma(\Theta^+(1540))$ is narrow due to a mixing between
 $([ud]^{6_c}_1\bar{s})$ and $[ud]^{3_c}_0\bar{s}$ such that the low mass eigenstate
 decouples from $KN$, then the $\vartheta^+$ should exist with a ``normal" width.
 With a mass even at 1700MeV the phase space ratio for $\vartheta \to K\bar{K}$ and
 $\Theta^+ \to KN$ is $\sim 16$. The $KK^*$ channel is open; the phase space
 enhancement in this case is  $\sim 5$ but the spin counting will elevate this so that
 we may expect a similar branching ratio to that of the $KK$ mode. The three body $KK\pi$
 mode will also contribute but uncorrelated three body modes are not expected to dominate over the 
two
 body ones. The net result is that we expect $\Gamma(\vartheta^+) \lesssim 100$MeV, such that the 
$\vartheta$ should be detectable  (likewise for the equivalent $S=-2$ member of the $\mathbf{10}$).

Surprisingly it is not
immediately possible to exclude such an exotic state in $K^+K^0$ if its
mass is $\sim 1.6 -1.7$ GeV, and a dedicated search is suggested in
e.g. $K^+N \to K^+K^0\Sigma$\cite{oldexpt}.
 Other exotic members of the multiplets are probably broad
and unobservable if the dynamics of ref\cite{vento} underpins the formation of
triquarks (see later).


 \subsection*{Other members of the meson $\mathbf{10\oplus\overline{10}}$}

 A unified convention for constructing the symmetry states for multiquarks is
 given in Table 1 of \cite{cd2}, reproduced here as Table 1. This gives the combinations of three labels
 for the symmetric and mixed ($\lambda$) states; the mixed ($\rho$) and antisymmetric
 follow trivially. The labels $A,B,C$ are defined 
 $A \equiv [ud] = \bar{s}; B \equiv [ds] = \bar{u}; C \equiv [su] = \bar{d}$; note that
  $[ud] \equiv (ud-du)/\sqrt{2}$, etc., the sign of the antisymmetric combination being important. 

\begin{table}
\hspace{-3.0cm} \begin{tabular}{l|c|c}
  & \10bar & ${\bf 8}_5$ \\
\hline
$\Theta^+$ & $AAA$ &\\
\hline
$p$ & $-(ACA + CAA +AAC)/\sqrt{3}$ & $-(ACA +CAA -2AAC)/\sqrt{6}$\\
$n$ & $(ABA +BAA +AAB)/\sqrt{3}$ & $(ABA + BAA -2AAB)/\sqrt{6}$\\
\hline
$\Sigma^+$ & $(CAC+ACC+CCA)/\sqrt{3}$ & $(CAC+ACC -2CCA)/\sqrt{6}$\\
$\Sigma^0$ & $-(ABC+BAC+ACB+CAB+BCA+CBA)/\sqrt{6}$ & $-(ABC+BAC +ACB+CAB-2BCA-2CBA)/\sqrt{12}$\\
$\Lambda^0$ & & $-(ABC-ACB+BAC-CAB)/2$\\
$\Sigma^-$ & $(BAB+ABB+BBA)/\sqrt{3}$ & $(BAB+ABB-2BBA)/\sqrt{6}$\\
\hline
$\Xi^+$ & $-CCC$ &\\
$\Xi^0$ & $(CBC+BCC+CCB)/\sqrt{3}$ & $(CBC+BCC-2CBB)/\sqrt{6}$\\
$\Xi^-$ & $-(CBB+BCB+BBC)/\sqrt{3}$ & $-(CBB+BCB -2 BBC)/\sqrt{6}$\\
$\Xi^{--}$ & $BBB$ &
\end{tabular}
\caption{Pentaquark wavefunctions where $ABC$ are defined in the text. Note that consistency requires the
meson octet to be defined with each $q\bar{q}$ positive except for $\pi^+ = -u\bar{d}$; $\bar{K}^0 = - s\bar{d}$ and then $\pi^0 = (u\bar{u} - d \bar{d})/\sqrt{2}$. In this convention $\eta_8 = (2s\bar{s} - u\bar{u} - d\bar{d})/\sqrt{6}$}
\end{table}

In the meson case it is useful to adopt the order  $A_1 B_2\equiv ([qq]\bar{q}),C_3 \equiv \bar{q}$    
and the $(..)$ shows which are understood to be in the triquark correlation. 

Triquarks  $[qq]\bar{q}$ are in flavour $\mathbf{\bar{3}\otimes \bar{3}=\bar{6}\oplus 3}$. To make 
a $\mathbf{\overline{10}\oplus{8}}$ of mesons (or pentaquarks) the triquark must be in flavour 
$\mathbf{\bar{6}}$, which is composed of the following members: \begin{center}
 $AA=[ud]\bar{s}$               \\
\vspace{1.5ex} 
 $\{AB\}=([ud]\bar{u}+[ds]\bar{s})/\sqrt{2}$\qquad\qquad 
$\{CA\}=([ud]\bar{d}+[su]\bar{s})/\sqrt{2}$\\
\vspace{1.5ex}
$BB=[ds]\bar{u}$\qquad\qquad\qquad $\{BC\}=([ds]\bar{d}+[su]\bar{u})/\sqrt{2}$\qquad\qquad\qquad 
$CC=[su]\bar{d}$\\
\end{center}

\par
The nonexotic combinations      in which $q_iq_j\bar{q_j}$ are in flavour $\mathbf{3}$  allow the 
possibility of $q_j\bar{q_j} \to $
  gluons and hence mixing with conventional hadrons; such combinations have no advantages in
  forming metastable states      and comprise the $P$-wave excited version of Jaffe's nonet 
\cite{scalardiq}. The combinations $\{AB\},\{BC\}$ and $\{CA\}$ are at least stable against  
$q_j\bar{q_j} \to $ gluons (by $SU(3)_F$) although in the mechanism of Vento \cite{vento}, it is 
only the $AA$ triquark that is fully stable, all others being unstable against decay into $\pi$ or 
$\eta_s$.

Using the conventions of \cite{cd2}, a full set of meson representations can be constructed:

\begin{tabular}{ll}
$\vartheta^+ = AAA$ & \\ 
$K^+_{\mathbf{\overline{10}}} = -(ACA+CAA+AAC)/\sqrt{3}$ & $K^+_{\mathbf{8b}} = -(ACA+CAA-2AAC)/\sqrt{6}$\\
$\pi^+_{\mathbf{\overline{10}}} = -(CAC+ACC+CCA)/\sqrt{3}$ & $\pi^+_{\mathbf{8b}} = -(CAC+ACC-2CCA)/\sqrt{6}$\\
$``K^+"= CCC$ & \\
\end{tabular}

The other charge combinations follow by acting on these with $I_-$ accordingly.
Charge conjugate analogues of these correspond to a  $\mathbf{10\oplus 8}$ .

 Ref.\cite{kl1} considered only the exotic states at the corner of the $\mathbf{\overline{10}}$.
The dynamics of ref.\cite{vento}, discussed earlier, suggest that
other configurations, such as $([su]\bar{d})$ (and  $([ds]\bar{u})$) are energetically unfavoured, 
 due to the presence of $[us]$ in place of
 $[ud]$, for which the instanton forces are less attractive, and unstable
 because $m([us]\bar{d}) \geq m(\pi) + m(s)$, which enables decay.  One consequence could appear to 
be that only the $\Theta^+$
 will be narrow in such a dynamics: the $\Xi^{+,--}$ contain
 $([su]\bar{d})$ or $([ds]\bar{u})$ which are unstable against decay into
 $\pi + s$, while the other states mix with {\bf 8}. 
  
  The absence of prominent signals  other than the $\Theta^+$
 in the \10bar, in particular the $\Sigma_5$\cite{fcqz03}, may thus be
 explained: if such correlations occur, then they only create
 metastable configurations if either $[ud]$ and/or $([ud]\bar{s})$ are 
 involved. 
 We argue that the same situation arises in the  $\mathbf{\overline{10}}$ and  $\mathbf{10}$ of 
mesons: the only exotic combinations with a chance of 
suppressed widths are the $\vartheta^+$ and $\vartheta^-$, containing the triquarks $AA$ and 
$\bar{A}\bar{A}$ respectively.

The exotic $S=-1$  $K^+$ $([su]\bar{d})\bar{d}$ and $K^{--}$  $([ds]\bar{u})\bar{u}$ contain 
triquarks
 that are  unstable against $\pi$ emission in the model of ref\cite{vento}. Thus we expect their 
widths to be at least 
300MeV.
 Identifying such states will be a challenge. 
 The mass gap between the $\vartheta^+$ and the $K^{+,0,-,--}$ states will be $\sim 2m(s - d) - 
m([us] - [ud])$. As the $m([us] - [ud])$ is likely
to be at least as large as $m(s-d)$  the
spread is likely to be only $\sim 100$ MeV.

The other combinations occur in nonexotic multiplets and in general will mix. Consider for example 
the $J^P =1^-$ states, $\rho^0$ or $\pi^0_1$. 
Depending on the mixing angle between the \10bar and {\bf 8 } basis states, the
mass eigenstates can be degenerate   
  or separated by up to $2(m(s)-m(d))$. In this latter case the physical
  states are the ideal mass eigenstates $[qq]\bar{q}\bar{q}$ and $[sq]\bar{q}\bar{s}$.
  This small mass range for the states suggests that the hidden flavour mass basis is more 
  representative than an $SU(3)_F$ multiplet basis.


Thus we need to count the number of states.
 For the neutral states with $I=1$ and $I=0$ there are six permutations of the distinct $ABC$
  labels. In the $SU(3)_F$ symmetry basis these correspond to
  
 \begin{tabular}{ll}
  \10bar: $I=1$; & {\bf 8}$^{\lambda}$:$(I=1) + (I=0)$;\\
  {\bf 8}$^{\rho}$:$(I=1) + (I=0)$; & {\bf 1}: $I=0$.
  \end{tabular}
  
  The latter trio correspond to the familiar $\rho; \omega_8; \omega_1$ combinations
  in the case of $1^-$ and as such are indistinguishable from radially excited \qq~ nonets.
  The former would correspond to a pair of $\rho$s and a single $\omega_8$, and hence would be 
novel.
  Thus counting the population of $I=1$ and $I=0$ vector mesons within an energy range of $\sim 
300$MeV
  can hint at which underlying multiplets are present. 
  
  While the $\rho(1700),\omega(1650),\phi(1680)$ form a candidate nonet, their masses are
  somewhat unnatural. The $\rho(1450)$ and $\omega(1420)$ are missing a partner to complete
  the set. Depending on whether this is $I=0$ or $I=1$ could be novel. The $K^*(1410)$ appears
  to be anomalously low in mass for \qq~ systems but fits naturally into the \10bar configuration.
  The widths of most of these states are hundreds of MeV.

The general feature is that for a given $J^{P(C)}$ six neutral members are expected within a few 
hundred MeV. If any of the plethora of observed states\cite{pdg04} is associated with
these, such that their widths of $\sim 300$MeV give a scale for their (in)stability, then a 
$\vartheta^+$ seems an unavoidable consequence.

\subsection*{Dynamics and $J^{PC}$ in $L=1$ $qq\bar{q}\bar{q}$ mesons}

An $L=1$ $qq\bar{q}\bar{q}$ system has a variety of $J^{PC}$ combinations. The relative masses
and potential stability of these can depend upon the correlations of strong QCD.
We now investigate the different dynamical arrangements for an $L=1$ $qq\bar{q}\bar{q}$ system, 
distinguished by the configuaration of the orbital angular momentum. We will see that dynamics may 
favour a triquark configuration for the  $\mathbf{18\oplus \overline{18}}$ of mesons, and that such 
a configuration has a limited spectum of $J^{PC}$ states. The same dynamical assumptions, namely 
the prevalence of ``good'' diquarks over ``bad'', explains the absence of higher representations 
such as $\mathbf{27}$.

In the diquark-diquark correlation, the  $qq$ and $\bar{q}\bar{q}$ systems are each in $L=0$ with 
an $L=1$ between them. For quark pairs which are symmetric in space, the remaining degrees of 
freedom must be antisymmetric. The allowed combinations are as follows, for total quark spin $S$,  
with $\{\quad \}$ and $[\quad ]$ denoting flavour  $\mathbf{6}$ ($\mathbf{\overline{6}}$) and  
$\mathbf{3}$ ($\mathbf{\overline{3}}$) respectively, and superscripts and subscripts denoting 
colour 
and spin:\\
\begin{tabular}{lccl}

&$\mathbf{ 10+8 }$                                      &       $\mathbf{ \overline{10}+8 }$                                    
& \\

(i)&${\{ qq \} }^{\bar{3}}_1{[\bar{q}\bar{q}]}^3_0$     &       ${[qq]}^{\bar{3}}_0  {\{ 
\bar{q}\bar{q} \} }^3_1 $       & $S=1,J^P=0^-,1^-,2^-$ \\
(ii)&${\{ qq \} }^6_0  {[\bar{q}\bar{q}]}^{\bar{6}}_1$   &       ${[qq]}^6_1      
  {\{ \bar{q}\bar{q} \} }^{\bar{6}}_0$& $S=1,J^P=0^-,1^-,2^-$ \\

\end{tabular}\par

A different set of configurations arises if the $L=1$ is \emph{between} a 
pair of quarks or antiquarks and the concept of a ``diquark'' dissolves. The quark pairs are now 
spatially antisymmetric, so that to satisfy the Pauli 
principle, the same flavour and colour correlations as above will have spin symmetry flipped
from 0 (antisymmetric) to 1 (symmetric), or vice versa. The resulting 
combinations, where $q|q$ denotes a pair of quarks in $L=1$, are:
\par
\begin{tabular}{lccl}

&$\mathbf{ 10+8 }$                                      &$\mathbf{ 
\overline{10}+8 }$                              & \\

(iii)&${\{ q|q \} }^{\bar{3}}_0  [{\bar{q}\bar{q}]}^3_0 $ &${[qq]}^{\bar{3}}_0  
{\{ \bar{q}|\bar{q} \} }^3_0 $  &$S=0,J^P=1^-$ \\
(iv)&${\{ q|q \} }^6_1   {[\bar{q}\bar{q}]}^{\bar{6}}_1$ &${[qq]}^6_1 {\{ 
\bar{q}|\bar{q} \} }^{\bar{6}}_1$       &$S=0,J^P=1^-$ \\
                                                        &                               
        &               &$S=1,J^P=0^-,1^-,2^-$ \\
                                                        &                               
        &               &$S=2,J^P=0^-,1^-,2^-,3^-$ \\
(v)&${\{ qq \} }^6_0  [{\bar{q}|\bar{q}]}^{\bar{6}}_0   $       &${[q|q]}^{\bar{3}}_1    
  {\{ \bar{q}\bar{q} \} }^3_1$ &$S=0,J^P=1^-$ \\
(vi)&${\{ qq \} }^{\bar{3}}_1    {[\bar{q}|\bar{q}]}^3_1$&${[q|q]}^6_0  {\{ 
\bar{q}\bar{q} \} }^{\bar{6}}_0 $       &$S=0,J^P=1^-$ \\

                                                        &                               
                &       &$S=1,J^P=0^-,1^-,2^-$ \\
                                                        &                               
                &       &$S=2,J^P=0^-,1^-,2^-,3^-$ \\

\\
\end{tabular}\par

The first thing to notice is that the two different pictures have different $J$ 
couplings. If the diquark-diquark picture describes the $\pi_1(1400;1600)$
mesons, then we would also expect $0^{-+}$ and $2^{-+}$ partners at comparable 
mass, shifted by spin orbit splittings. Conversely, the latter picture 
allows for $S=0,1$ or $2$. If we supposed that the dynamics were such that 
the $S=0$ state was favoured (and we will argue that this could be so), then the apparent absence 
of $0^{-+}$ and $2^{-+}$ 
siblings to the $\pi_1$ mesons is natural.

 In the diquark-diquark picture we see that a meson in 
$\mathbf{18}$ or $\mathbf{\overline{18}}$ is made of a ``good'' 
and ``bad'' diquark: 
in Jaffe's original paper\cite{scalardiq}, the absence of $S$-wave mesons in this flavour 
representation was due to these repulsive colourmagnetic interactions. By contrast, we see that in 
the second picture, in which the  $L=1$ is ``within'' a diquark, the spatial antisymmetry annuls 
these repulsive forces, turning
 a ``bad'' diquark of a given flavour (colour-spin symmetric) into a ``good'' 
diquark (colour-spin symmetric), as in configurations (iii) and (iv) (or conversely turns a ``good" 
diquark into a ``bad'' diquark in configurations (v) and (vi)). However, 
once the orbital angular momentum separates quarks, the short-range 
hyperfine interaction is heavily suppressed, so it is better to describe these
 $L=1$ diquarks as neither ``good'', nor ``bad'', but ``null''. 

What then can we say about the dynamics of a $[qq]\{\bar{q}|\bar{q}\}$ system? On the one hand, we 
could consider this sytem a direct mirror image of the JW correlation\cite{jw} for the $\Theta^+$, 
$(ud)\ |\ (ud)\quad \bar{s} \to \bar{q}\ |\ \bar{q}\quad [qq]$. 
However, the $[q q]^{\bar{3}}_0\bar{q}$ and $[q q] ^6_1 \bar{q}$ 
  systems in $S=1/2$ are very light due to attractive colourmagnetic and instanton forces 
\cite{hog}\cite{vento}, so it is natural to consider the configurations  (iii)  and  (iv)  as a 
tightly bound $S$-wave $qq\bar{q}$ triquark in a 
  relative $P$-wave with a $\bar{q}$, (along with their charge conjugates)\footnote{ 
  Provided the triquark is compact spatially, the difference between a $P$-wave between 
  the antiquarks and a $P$-wave between the free antiquark and the triquark can be neglected}. Each 
of these systems has a ``good'' diquark,
  and the presence of a $\bar{q}$ in $S$-wave with these diquarks lowers 
  the energy further. 

Refs\cite{hog} and \cite{maltman} noted that the colourmagnetic interaction 
mixes the $S=1/2$ $[q q]^{\bar{3}}_0\bar{q}$ and $[q q] ^6_1 \bar{q}$ triquarks (configurations 
(iii) and (iv)) to give 
the lowest eigenvalue $-21.88C$ in the case of full  $SU(3)_F$ symmetry, a considerably stronger 
attraction than the diquark-diquark system, for which the lightest configuration has 
an energy shift of $-16/3C$.
Ref.\cite{vento} noted that instanton forces cause the same mixing and lower the energy
of the triquark correlation further.

Note also that the mixing of the (iii) and (iv) configurations, with their
downward energy shifts, occur only in the $S=0, J^P=1^-$ state. Thus if this
correlation is dominant, the lowest lying multiplets can be those with $J^P =1^-$, all other $J^P$ 
states being higher 
in energy. In any event, this possible dynamical picture limits the spectrum of $J^{PC}$ that 
could be expected, so that this picture is less readily falsified by the dearth of experimental 
evidence of mesons in  $\mathbf{10}$ or $\mathbf{\overline{10}}$.

The same dynamics suggests that higher representations such as $\mathbf{6\otimes 6=27\oplus 8\oplus 
1}$ are not dynamically favoured. $\{qq\}\{\bar{q}\bar{q}\}$ systems are made of pairs of ``bad'' 
diquarks, and even moving to the triquark picture cannot annul these repulsive forces.

Thus we suggest that although tetraquark states in general will tend to be broad, the $J^P = 1^-$ 
\10bar and {\bf 10}, possibly mixed with {\bf 8}, may have the best chance of being observable. 
There are subtleties involved with
forming the $C = \pm$ eigenstates associated with  $\mathbf{18\oplus \overline{18}}$; these are
discussed
in an appendix.   We shall see later that the multiquark dynamics leading to the above also imply 
that there
  exists a ``mirror" multiplet with $J^{PC} = 1^{--}$ neutral members.

The phenomonology of the  controversial $\pi_1(1400)$ and $\pi_1(1600)$ is not
inconsistent with the triquark correlation. The dynamically preferred arrangement in triquark 
language has $S=0$, which fits in neatly with the absence of clear $0^-$ and $2^-$ partners to the 
$\pi_1(1400)$ and $\pi_1(1600)$. It is plausible that the mixed $[q q]^{\bar{3}}_0\bar{q}$ and $[q 
q] ^6_1 \bar{q}$ system could be the only stable triquark correlation, and that states which do not 
benefit from this mixing ($S=1,2$) are not dynamically preferred.


\subsection*{$J^{PC} = 1^{-\pm}$ multiplets in correlated quark models}

 In the search for evidence of gluonic degrees of freedom in Strong QCD, attention
 has focussed in particular on the prediction of exotic  quantum numbers  such as $J^{PC} = 
1^{-+}$,
 which are forbidden for \qq~  in a potential but allowed if the gluonic degrees of freedom are
 excited. The masses of the lightest such hybrids are predicted from lattice QCD
 to be above 1.8GeV\cite{lattice}. Thus
 the appearance of $\pi_1(1600) \to \pi \eta'$ is intriguing as this was long ago
 suggested as a selection rule by Lipkin (see
 citation to unpublished remarks by Lipkin in \cite{lee}) and then in a modern context
 in\cite{fclipkin1} where the decay of a hybrid $1^{-+}$ was predicted to have $\pi\eta' > 
\pi\eta$. A problem is that such a state could also occur from $qq\bar{q}\bar{q}$ and the mass 
pattern
 of the other members of
 the nonet would need to be identified in order to distinguish this from a canonical
 $q\bar{q}g$ hybrid.  In this context there is also reported a companion state $\pi_1(1400)$, which 
is seen in $\pi \eta$
 but not $\pi\eta'$.  If such a decay
 conserves flavour, and if $\eta \equiv \eta_{\mathbf{8}}$, then such a state cannot belong to an {\bf 8}. 
 Motivated by this state, ref\cite{chung} proposed that $\pi_1(1400)$ belong to
 a {\bf 10} $\oplus$ \10bar $ qq\bar{q}\bar{q}$ multiplet, and thus is not
 a hybrid \qq$g$ state in {\bf 8}.

As shown in the appendix, we can take a linear combinations of $\pi^+_{\mathbf{10}}$ and $\pi^+_{\mathbf{\overline{10}}}$ 
to give a $I^G(J^{PC}) = 1^-(1^{-+})$ state overlapping to $\pi^+\eta_{\mathbf{8}}$, and likewise a 
combinations of  $\pi^+_{\mathbf{8_a}}$ and $\pi^+_{\mathbf{8_b}}$ overlapping to  $\pi^+\eta_{\mathbf{1}}$. The combinations 
$CAC$ and $CCA$ (and charge conjugates) have 2 strange masses, while $ACC$ has 0 strange masses, so 
we can express the $SU(3)_F$ symmetry basis states $X$ and $Y$ for the  $1^-(1^{-+})$ states as:
\bea
X=(\pi^+_{\mathbf{10}}\oplus\pi^+_{\mathbf{\overline{10}}})/\sqrt{2}&=&[\pi^0\eta_{\mathbf{8}}]=(\sqrt{2}qs\bar{q}\bar{s}+
qq\bar{q}\bar{q})/\sqrt{3} \nonumber\\
Y=(\pi^+_{\mathbf{8_a}}\oplus\pi^+_{\mathbf{8_b}})/\sqrt{2}&=&[\pi^0\eta_{\mathbf{1}}]=(-\sqrt{2}qs\bar{q}\bar{s}+2qq\bar{q}\bar
{q})/\sqrt{6} \nonumber
\eea
where $qs\bar{q}\bar{s}$ and $qq\bar{q}\bar{q}$ denote the (normalised) parts of the wavefunction 
with 2 and 0 strange quarks respectively. The orthogonal linear combinations give an 
$I^G(J^{PC}) = 1^+(1^{--})$ $\rho$ state overlapping to meson states $K\overline{K}$ and $\pi\pi$. 
For ref.\cite{chung}, who assume $\eta \equiv \eta_{\mathbf{8}}$ and $\eta' \equiv \eta_\mathbf{1}$, it is 
states $X,Y$ 
respectively that decay to $\pi\eta$ and $\pi\eta'$:
the physical $\eta$ and $\eta'$ are not pure {\bf 8} and {\bf 1}. One could choose the mixing of 
the decuplet and octet states  $X$ and $Y$
to enforce decays to the $\pi\eta$ and $\pi\eta'$ respectively; this would require the mixing angle
of $X,Y$ to be the same as the $\eta-\eta'$ mixing.  The eigenstates mixed in this way would then 
correspond to $\pi_1(1400)\to \eta \pi$ and  $\pi_1(1600)\to \eta' \pi$. Conversely, mass 
eigenstates that are ideal
 
\bea
X_H=qq\bar{q}\bar{q}\nonumber\\
X_L=qs\bar{q}\bar{s}\nonumber
\eea
will decay to $\pi\eta_n$ and $\pi\eta_s$ respectively.

In order to go from mass eigenstates to symmetry basis states, 
it is necessary to
have a mixing amplitude $A$[(\dd$u)\to$ (\ss$u$)]  $>m$(\dd$u)- m$(\ss$u$). The stability
of the \ss$u$ in contrast to the \dd$u$ seems to argue against that but a definite answer is 
beyond our ability to determine without further assumptions. If $\pi_1(1400;1600)$ are 
identified with these states, then the  quark mass eigenstates appear to be nearly realised: the 
masses based on the simplest flavour
counting are consistent with $\pi^+_1(1400)$ and $\pi^+_1(1600)$ as  the (dominantly)  
$qq\bar{q}\bar{q}$
and $qs\bar{q}\bar{s}$ states. 
This agrees with our earlier prediction that $m(\vartheta)$, the \udss~ state is also $ \sim 1600$MeV,
which in turn is consistent with the $\Theta^+(1540)$. A pair of $\rho$ states with similar
masses to those of the two $\pi_1$s are required. There are known problems with identifying the
$\rho(1460;1600)$ as simply \qq~ states\cite{cdk} and the existence of $\rho(1250)$
remains uncertain. If the latter exists as a \qq~ candidate, then the other pair may be related to 
the  $qq\bar{q}\bar{q}$ and $qs\bar{q}\bar{s}$ states. Conversely, if the $\rho(1460)$ is the 
lightest such resonance, then the
$K^*(1410)$ mass is more in tune with the pattern of interest here than it would
be for a \qq~ nonet. A test will be the presence or absence of isoscalar partners to these 
states. A {\bf 10} or \10bar have no such $I=0$ $\eta_1$ or $\omega$ states whereas the {\bf 8}
does; a canonical nonet would have the $\rho; \omega; \phi$ analogues. 

 In any case, the $\sim 300$MeV width of the $\pi_1(1400)$ and $\pi_1(1600)$ is consistent with a 
triquark dynamics. The wavefunction of a $\pi$ state, in any decuplet-octet mixing scenario, is 
composed of triquarks that can decay to $\pi+s, \pi+q$ or $\eta_s+q$ (or charge conjugates, for the 
antitriquark component).

In \cite{chung} it is shown that if the $1^{-+}$ $\pi_1(1400) \to \eta\pi$ is a member of a
\10bar $\oplus$ {\bf 10} then there must also be a partner state $\rho \to \pi\pi$ and $K\bar{K}$.
 Using symmetry arguments, ref.\cite{chung} show that this ``supermultiplet'' of  
$\mathbf{10\oplus\overline{10}}$ decaying to two pseudoscalar mesons (PP) ought to be accompanied 
by another supermultiplet decaying to a pseudoscalar and vector meson (PV), a $I^G(J^{PC}) = 
1^+(1^{--})\to \omega\pi$ and a $1^-(1^{-+})\to \rho\pi,K^* \bar{K}$. Note, however, that the 
$1^{-\pm}$ in the PV supermultiplet mentioned by CKK in ref.\cite{chung}  must be accompanied by 
siblings $0^{-\pm}$ and $2^{-\pm}$. By visualising the system as a triquark-antiquark in a 
$P$-wave, we can immediately understand the origin of the two supermultiplets and their $J^{PC}$ 
quantum numbers.

In normal $q\bar{q}$ mesons, $q$ and $\bar{q}$ have opposite parity, so that $P$-wave states have 
$P=+$. Conversely, in the triquark-antiquark picture, the triquark $Q$ and antiquark $\bar{q}$ have 
the $same$ parity, so that $P$-wave states have $P=-$. In the appendix, we show that the 
wavefunction of a $qq\bar{q}\bar{q}$ meson has an extra degree of freedom compared to a $q\bar{q}$ 
meson,  manifested in the freedom to take $\mathbf{10+\overline{10}}$ versus 
$\mathbf{10-\overline{10}}$ (or equivalently $\mathbf{8_a+8_b}$ versus $\mathbf{8_a-8_b}$), giving 
$C=+$ and $C=-$ for each $J$. So if we take the $q\bar{q}$ spectrum of states,
\bea
        S=0, J=1^{+-}   & &      S=1,J=0^{++},1^{++},2^{++}
\eea
flip the parity, and take  $C=\pm$, we acquire precisely the spectrum of \cite{chung}:
\bea
S=0, J=1^{--}   & &      S=1,J=0^{--},1^{--},2^{--}\\
 S=0, J=1^{-+}  & &      S=1,J=0^{-+},1^{-+},2^{-+}
\eea

 In the work of Chung et al \cite{chung} there is no dynamical picture that distinguishes the two 
supermultiplets, and hence no suggestion 
as to why the PV supermultiplet has not been experimentally observed. We show
in the appendix that there are different dynamics underpinning the supermultiplets. The PP 
supermultiplet, to which the observed $\pi_1\to \eta\pi$ belongs, has $S=0$, and we saw earlier 
that this spin configuration is the only one in which  mixing from one gluon exchange and instanton 
forces allows a light and metastable triquark. On the contrary, the PV multiplet has $S=1$, a 
configuration in which mixing cannot occur, resulting in heavier, possibly unstable triquarks. This  
might account for the experimental elusiveness of the PV supermultiplet.

Earlier we noted that the diquark-diquark picture can only have $S=1, J=0^-,1^-,2^-$. In the 
appendix we show that for the diquark-diquark configuration $\zeta=(-1)^l$, so that such a system 
in $P$-wave can exist in only $one$ supermultiplet $(\zeta=-1)$. Due to  angular momentum 
conservation in the decays of the
 $0^-$ and $2^-$, this supermultiplet can only be PV. Thus, provided it is valid to treat diquarks 
as effective bosons, the $1^{-+}$ $\pi_1(1400)\to\eta\pi$ and $\pi_1(1600)\to\eta'\pi$ cannot be in 
the diquark-diquark arrangement.


\subsection*{Conclusion}

In conclusion, confirmation of a narrow $\Theta^+(1540)$ and the absence of other narrow members of 
a \10bar can be explained by correlations
that suggest there should be a $\vartheta$ with canonical width $\lesssim 100$MeV
together with
a family of broad partners. This particular dynamics is exceptional as 
received wisdom has been that all these states should ``fall-apart".  If the $\Theta^+$ is an 
artefact, or if its narrow width 
is due to  some mechanism other than the mixing among correlations as discussed here, then the 
$1^{-(\pm)}$ tetraquark mesons will be all broad as in \cite{jaf77,jafexotic}. In any event, if
the existence of a resonant $\Theta^+$ with narrow width survives further scrutiny, then
a \10bar of mesons with moderate widths, of which the $\vartheta$ may have a canonical width,
merits investigation.

Models which consider four-quark mesons produce a considerable multiplicity of flavour and spin 
states. Using a triquark-quark correlation
for tetraquarks there can be a exception to this rule, with only a reduced set of states appearing
and the possibility that among these the $J^P =1^-$ \10bar and {\bf 10} may contain
observable states. It seems possible to associate certain otherwise peculiar states in the meson 
spectrum with those predicted here.

This model is trivially falsifiable by comparision of its prediction of flavour exotic states with 
experiment. We have noted that while the $\rho(1700),\omega(1650),\phi(1680)$ form a candidate 
nonet, their masses are
  somewhat unnatural. The $\rho(1450)$ and $\omega(1420)$ are missing a partner to complete
  the set and determine whether they are in a nonet or \10bar. The $K^*(1410)$ appears
  to be anomalously low in mass for \qq~ systems but fits naturally into the \10bar configuration.
  The widths of most of these states are hundreds of MeV.

The general feature is that for a given $J^{P(C)}$ six neutral members are expected within a few 
hundred MeV. If any of the plethora of observed states\cite{pdg04} is associated with
these, such that their widths of $\sim 300$MeV give a scale for their (in)stability, then a 
$\vartheta^+$ with a canonical width seems an unavoidable consequence.


 \subsection*{Appendix}

To obtain charge conjugation eigenstates for tetraquark mesons we need to write their wavefunctions 
in $qq\bar{q}\bar{q}$ and $\bar{q}\bar{q}qq$ form. We demonstrate the procedure for the nonstange 
neutral members of the $\mathbf{10}$ and $\mathbf{\overline{10}}$, noting that we can easily 
generalise to the $I_3=\pm 1$ members with the usual $G$ parity operator. Note also that the 
following analyses work for the  $\mathbf{8_a}$ and  $\mathbf{8_b}$.

From the wavefunctions given earlier , we find
\bea
\pi_{\mathbf{10}}^0=\mathbf{6\otimes 3}&=&-( \{ds\}[\bar{d}\bar{s}]
+\{su\}[\bar{s}\bar{u}]
+\{ud\}[\bar{u}\bar{d}])/\sqrt{3} \\
\widetilde{\pi}_{\mathbf{10}}^0=\mathbf{3\otimes 6}&=&-
([\bar{d}\bar{s}]\{ds\}
+[\bar{s}\bar{u}]\{su\}
+[\bar{u}\bar{d}]\{ud\})/\sqrt{3} \\
\pi_{\mathbf{\overline{10}}}^0=\mathbf{\bar{3}\otimes\bar{6}}&=&-
([ds]\{\bar{d}\bar{s}\}
+[su]\{\bar{s}\bar{u}\}
+[ud]\{\bar{u}\bar{d}\})/\sqrt{3}\\
\widetilde{\pi}_{\mathbf{\overline{10}}}^0=\mathbf{\bar{6}\otimes \bar{3}}&=&-
(\{\bar{d}\bar{s}\}[ds]
+\{\bar{s}\bar{u}\}[su]
+\{\bar{u}\bar{d}\}[ud])/\sqrt{3}
\eea
Notice $C\pi_{\mathbf{10}}^0=\widetilde{\pi}_{\mathbf{\overline{10}}}^0$ and  
$C\pi_{\mathbf{\overline{10}}}^0=\widetilde{\pi}_{\mathbf{10}}^0$, so if we denote the 
$\mathbf{10}$ and $\mathbf{\overline{10}}$ wavefunctions
\bea
 \chi(\mathbf{10},\zeta)=\pi_{\mathbf{10}}^0+\zeta\widetilde{\pi}_{\mathbf{10}}^0 & \qquad &
\chi(\mathbf{\overline{10}},\zeta)=\pi_{\mathbf{\overline{10}}}^0+
\zeta\widetilde{\pi}_{\mathbf{\overline{10}}}^0
\eea
then our $\zeta$ is precisely that defined by equation (2) in \cite{chung}:\bea
C\chi(\mathbf{10},\zeta)=\zeta\chi(\mathbf{\overline{10}},\zeta)&\qquad&
C\chi(\mathbf{\overline{10}},\zeta)=\zeta\chi(\mathbf{10},\zeta)
\eea
We have the freedom to choose the phase between the $\mathbf{10}$ and $\overline{\mathbf{10}}$ , 
which we denote by $a$, so that the full
 wavefunctions of a four quark $\mathbf{10\oplus\overline{10}}$ have two degrees of freedom, $a$ 
and $\zeta$,
\bea
\pi^0(\zeta,a)&=&\pi_{\mathbf{10}}^0+\zeta\widetilde{\pi}_{\mathbf{10}}^0
+a\left(\pi_{\mathbf{\overline{10}}}^0+\zeta\widetilde{\pi}_{\mathbf{\overline{10}}}^0)\right)\eea
We see that
 $C\pi^0(\zeta,a)=\zeta a\pi^0(\zeta,a)$
 and the doubling of states follows: in the $\zeta=-1$ multiplet we have a $1^{--}$  $(\zeta a=-+)$ 
and $1^{-+}$ $(\zeta a=--)$, and in the $\zeta=+1$ multiplet we have a $1^{--}$  $(\zeta a=+-)$ and 
$1^{-+}$ $(\zeta a=++)$. Looking at the $qq\bar{q}\bar{q}$ part of the wavefunction only, we see 
that under interchange of quarks $2\leftrightarrow 3$ linear combinations of $\pi_{\mathbf{10}}^0$ 
and $\pi_{\mathbf{\overline{10}}}^0$ correspond to CKK's flavour wavefunctions in 
$q\bar{q}q\bar{q}$ form:  the $1^{-+}$ state is 
$\pi_{\mathbf{10}}^0+\pi_{\mathbf{\overline{10}}}^0\equiv [\pi^0\eta_8]$, the  $1^{--}$ state is 
$\pi_{\mathbf{10}}^0-\pi_{\mathbf{\overline{10}}}^0\equiv\sqrt{1\over 
3}\left([\pi^-\pi^+]+[\overline{K^0}K^0]+[K^-K^+] \right)$ (our $\pi^+$ is defined as 
$-u\bar{d}$\cite{cd2}, hence the phase difference compared to CKK). Likewise for linear 
combinations of the wavefunctions $\mathbf{8_a}$ and $\mathbf{8_b}$.

We can think of a $P$-wave tetraquark meson as a system of two quasi particles in $L=1$: in the 
diquark-diquark picture, two bosons of spin 0 and 1, in the triquark picture, two fermions of spin 
1/2 (we assume the lightest triquark is spin 1/2 following \cite{chung}\cite{vento}).
 In the JW picture of the $\Theta^+$ \cite{jw}, the $(ud)$ pairs are treated as effective bosons so 
that bose statistics forces a $P$-wave. However, this picture suffers from the fact that the $uudd$ 
wavefunction is not fully antisymmetrised, only the $(ud)$ pairs individually. In the meson world 
there is no such problem. We can treat a $qq\bar{q}\bar{q}$ system as a pair of bosons or pair of 
fermions at the same time ensuring their fermionic wavefunctions are fully antisymmetrised.

\par
Let us look firstly at the diquark-diquark correlation. Denoting $\{qq\}$ by $b$ and $[qq]$ by $B$, 
in shorthand notation we can write $\pi_{\mathbf{10}}^0=b\bar{B}$,
$\widetilde{\pi}_{\mathbf{10}}^0=\bar{B}b$,
$\pi_{\mathbf{\overline{10}}}^0=B\bar{b}$,
$\widetilde{\pi}_{\mathbf{\overline{10}}}^0=\bar{b}B$,so that the full wavefunctions can be 
expressed
\bea
\pi^0(\zeta,a)&=&b_1\bar{B}_2+\zeta \bar{B_1}b_2
+a\left(B_1\bar{b_2}+\zeta\bar{b_1}B_2\right)\eea
where the labels 1 and 2 denote the spin and space degrees of freedom of the bosons. 
$C\pi^0(\zeta,a)=\zeta a \pi^0(\zeta,a)$ in this shorthand notation, as expected. Interchanging the 
space and spin labels in the wavefunction brings a factor $(-1)^l$
\bea
\pi^0(\zeta,a) & = &(-1)^l[b_2\bar{B}_1+\zeta \bar{B_2}b_1
+a(B_2\bar{b_1}+\zeta\bar{b_2}B_1)]
\eea
and since bosons are commuting variables ($[b_1,\bar{B}_2]=0$),
\bea
\pi^0(\zeta,a) & = &(-1)^l[\bar{B}_1 b_2+\zeta b_1 \bar{B_2}
+a(\bar{b_1}B_2+\zeta B_1\bar{b_2})]\\
&=&\zeta(-1)^l\pi_1^0(\zeta,a)
\eea
so in the diquark-diquark picture $\zeta=(-1)^l$.
\par

In the triquark-antiquark  configuration there are two spin 1/2 fermions in $L=1$. Denoting the 
triquarks $\{AB\}$, $\{CA\}$ and $\{BC\}$ by $U,D$ and $S$, 
\bea
\pi_{\mathbf{10}}^0&=&(-u\bar{U}+d\bar{D}+s\bar{S})/\sqrt{3}=q\bar{Q} \\
\widetilde{\pi}_{\mathbf{10}}^0&=&(-\bar{U}u+\bar{D}d+\bar{S}s)/\sqrt{3}=\bar{Q}q \\
\pi_{\mathbf{\overline{10}}}^0&=&(-U\bar{u}+D\bar{d}+S\bar{s})/\sqrt{3}=Q\bar{q} \\
\widetilde{\pi}_{\mathbf{\overline{10}}}^0&=&(-\bar{u}U+\bar{d}D+\bar{s}S)/\sqrt{3}=\bar{q}Q 
\eea
and the full wavefunction is
\bea
\pi^0(\zeta,a)&=&q_1\bar{Q}_2+\zeta \bar{Q_1}q_2
+a\left(Q_1\bar{q_2}+\zeta\bar{q_1}Q_2\right)\\
&=& (-1)^{s+1}(-1)^l [  q_2\bar{Q}_1+\zeta \bar{Q_2}q_1
+a\left(Q_2\bar{q_1}+\zeta\bar{q_2}Q_1\right) ]\\
&=&  (-1)^{l+s} [  \bar{Q}_1q_2+\zeta q_1\bar{Q_2}
+a\left(\bar{q_1}Q_2+\zeta Q_1\bar{q_2}\right) ]\\
&=&(-1)^{l+s}\zeta\pi_1^0(\zeta,a)
\eea
since interchanging the labels $1$ and $2$ brings factors $(-1)^{s+1}$
and $(-1)^l$ for the spin and space labels respectively, and for fermions 
$\{q_1,\bar{Q}_2\}=0$. Thus in the triquark antiquark picture $\zeta=(-1)^{l+s}$. Once again 
$C\pi^0(\zeta,a)=\zeta a \pi^0(\zeta,a)$ in this shorthand notation, as expected.
\par

In the work of
Chung et al \cite{chung}, a quantum number $\zeta$ distinguishes the two supermultiplets, being 
labelled by $\zeta=+1$ (PP) and $\zeta=-1$ (PV), or vice versa (the overall phase of $\zeta$ is not 
important). At first glance, it appears as though the  total spin $S$ of the quarks distinguishes 
the two supermultiplets: the $S=0$ states being PP and the $S=1$ states being PV, but some caution 
is needed. By  angular momentum conservation for $P$-wave decays, 
$0^-,2^-\to$ PV but
not to PP,
so we can immediately assign the $0^-$ and $2^-$ states to the PV multiplet. However,
since we can have both
$1^-\to$ PV  and 
$\to$ PP in P wave, there is nothing, \emph{a priori}, that tells us to which supermultiplet a 
$1^-$ must belong. In the appendix, though, we use Fermi-Dirac statistics to show that for a 
$P$-wave triquark-antiquark system, the supermultiplet label $\zeta$ is given by 
$\zeta=(-1)^{s+1}$. We can confirm, then, what we had already suspected: that the total spin $S$ of 
the quarks determines precisely to which supermultiplet a state belongs. The states with
$S=0$ belong to the PP supermultiplet ($\zeta=-1$), and those with $S=1$ to the PV supermultiplet 
($\zeta=+1$).

As a brief digression, it's worth considering  the $P$-wave excited version of Jaffe's original 
nonet in the triquark-antiquark correlation.
In this case, we must have a flavour $\mathbf{3}$ of triquarks, where $SU(3)_F$ does not protect 
the annihilation of $q\bar q$ pairs, so stable triquarks cannot form. Annihilation of $q_i 
q\bar{q}\to q_i$ is equivalent to rewriting $U\to u$, $D\to d$ and  $S\to s$, or simply $Q\to q$ in 
our shorthand notation. In that case, the wavefunction is clearly zero for $a=-1$. The extra degree 
of freedom has been stripped away, and we recover the ordinary spectrum of $q\bar{q}$ states:
\bea
\zeta a = -+ & & S=0, J=1^{+-}\\
\zeta a = ++ & & S=1, J=0^{++},1^{++},2^{++}
\eea
Thus in the triquark-antiquark picture, a nonet of $P=-$ tetraquarks will not form. On the 
contrary, in the diquark-diquark picture the $q\bar{q}$ annihilation is inhibited by the $P$-wave 
barrier so it is possible that a nonet should be seen.

\bc {\bf Acknowledgments} \ec
  
This work is supported, in part, by grants from the Particle Physics
and Astronomy Research Council, the EU-TMR program ``Euridice''
HPRN-CT-2002-00311, a Clarendon Fund Bursary and by the U.S. Department of Energy under
contract DE-AC05-84ER40150.




\begin{thebibliography}{99}  

\bibitem{scalardiq} R.L.Jaffe, Phys.Rev. D {\bf 15} 267 (1977);
  F.E.Close and N.Tornqvist, J.Phys.G. {\bf 28} R249 (2002)
\bibitem{dgg} A.De Rujula, H.Georgi and S.L.Glashow, Phys. Rev. D{\bf
    12}, 147 (1975);Ya.B. Zeldovich and A.D. Sakharov, Yad. Fiz {\bf
    4}(1966)395; Sov. J. Nucl. Phys. {\bf 4}(1967)283.
 \bibitem{jw} R.L.Jaffe and F.Wilczek, hep-ph/0307341
\bibitem{kl1} M.Karliner and H.J.Lipkin, hep-ph/0307243
 \bibitem{jafexotic} R.L. Jaffe, hep-ph/0409065
\bibitem{maltman} B.Jennings and K.Maltman, hep-ph/0308286
\bibitem{hog} H.Hogaasen and P.Sorba, hep-ph/0406078
 \bibitem{vento} N.I.Kochelev, H.J.Lee and V.Vento, hep-ph/0404065 
\bibitem{pdg04} Particle Data Group, Phys Letters B592 (2004) 1
\bibitem{jaf77} R.L.Jaffe, Phys.Rev. D {\bf 15} 267 (1977)
\bibitem{ct02} F.E.Close and N.Tornqvist, J.Phys.G. {\bf 28} R249 (2002)
\bibitem{maiani} L.Maiani, F.Piccinini, A.D.Polosa, V.Riquer,  hep-ph/0407017 
\bibitem{chung} S.U.Chung, E.Klempt and J G Korner, hep-ph/0211100; S.U.Chung and E.Klempt,
hep-ph/0306018
\bibitem{indiana}  A.~R.~Dzierba {\it et al.}, Phys.\ Rev.\ D {\bf 67} (2003) 094015
\bibitem{fc04ichep} F.E.Close, Rapporteur talk at ICHEP04
\bibitem{klmix} M.Karliner and H.J.Lipkin, Phys.Lett.B {\bf 586} 303, 2004 
\bibitem{jjdFermilab} J.J.Dudek talk at the First Meeting of the APS Topical Group on Hadronic Physics, Fermilab November 2004 (Proceedings to appear in J.Phys.G)
\bibitem{fcdudekls} F.E.Close and J.J.Dudek, Phys.Lett.B {\bf 583}
  278, 2004 
\bibitem{oldexpt} W.P.Dodd, T.Joldersma, R.B.Palmer and N.P.Samios, Phys.\ Rev.\  {\bf 177} (1969) 
1991.
\bibitem{cd2} F.E.Close and J J Dudek, hep-ph/0401192
\bibitem{fcqz03} F.E.Close and Qiang Zhao, hep-ph/0404075 
\bibitem{lattice}   
P. Lacock, C. Michael, P. Boyle and P. Rowland, Phys.Lett. B {\bf 401}, 308   
(1997);   
C. Bernard {\it et al.}, Phys.Rev. D {\bf 56}, 7039 (1997);   
P. Lacock and K. Schilling, Nucl. Phys. Proc. Suppl. {\bf 73}, 261 (1999);   
C. McNeile, hep-lat/9904013;   
C. Morningstar, Nucl. Phys. Proc. Suppl. {\bf 90}, 214 (2000)   
\bibitem{lee} B W Lee, S Okubo and J Schecter Phys Rev. 135 (1964) B219
\bibitem{fclipkin1} F.E.Close and H.J.Lipkin,Phys Lett 196 (1987) 245;















\bibitem{cdk}   
F.E. Close, A. Donnachie and Yu.S. Kalashnikova, Phys.Rev. D {\bf 65},   
092003 (2002)  
  
\end{thebibliography}
\end{document}